\newif\ifdraft
\draftfalse
\newif\ifarxiv
\arxivfalse

\newif\ifeasychair
\easychairtrue

\ifdraft
\documentclass[draft]{llncs}
\else
\documentclass{llncs}
\fi

\newcommand{\query}[1]{\marginnote{\raggedright\footnotesize\itshape\hrule\smallskip{#1}\smallskip\hrule}}
\renewcommand{\query}[1]{} % Uncomment to turn off marginnotes

\newcommand{\M}{Mizar}
\newcommand{\mml}{MML}

\newif\iffinal
\finaltrue
\newif\ifpublic\publictrue

\usepackage{savesym}
\usepackage{calc}

\usepackage[T1]{fontenc}
\usepackage[utf8]{inputenc}
\usepackage{lmodern}
\usepackage[scaled=.8]{beramono}
\usepackage{latexsym}
\usepackage{tabularx}
\usepackage{varwidth}
\usepackage{textcomp}
\savesymbol{iint}
\savesymbol{iiint}
\usepackage{wasysym}
\restoresymbol{wasy}{iint}
\restoresymbol{wasy}{iiint}
\usepackage{amssymb}
\usepackage{unicode-chars}
\ifdraft\else
\fi

\usepackage[british]{babel}
\usepackage[final]{graphicx}
\usepackage{wrapfig}
\usepackage[inline]{enumitem}
\usepackage{ctable}
\usepackage{multirow}
\usepackage{booktabs}
\usepackage[babel]{csquotes}
\MakeAutoQuote{“}{”}
\MakeAutoQuote*{‘}{’}
\usepackage{bm}
\usepackage{mathtools}
\usepackage{lstsemantic}
\usepackage{lstlangmizar}
\usepackage{soul}
\setul{.25ex}{}
\usepackage{ragged2e}

\usepackage{modules}

\usepackage[obeyDraft,textsize=tiny]{todonotes}

\usepackage[backend=bibtex,hyperref=auto,
firstinits=true,style=numeric,
urldate=iso8601,isbn=false,doi=false]{biblatex}
\ifeasychair
\else
\fi
\DeclareNameAlias{author}{last-first}
\DeclareNameAlias{editor}{last-first}
\DeclareNameAlias{translator}{last-first}
\DeclareFieldFormat{labelnumberwidth}{#1.}
\DeclareFieldFormat{title}{#1}
\DeclareFieldFormat
  [article,inbook,incollection,inproceedings,patent,thesis,unpublished]
  {title}{#1}
\DeclareFieldFormat{journaltitle}{#1}
\newbibmacro*{volume+number+eid}{%
  \printfield{volume}%
  \iffieldundef{number}
    {}
    {(%
      \printfield{number}%
     )%
    }%
  \setunit{\addcomma\space}%
  \printfield{eid}}
\DeclareFieldFormat{url}{\url{#1}}

\usepackage{url}

\begin{document}
\title{A Qualitative Comparison of\iffinal\\ \fi the Suitability of Four Theorem Provers\iffinal\\ \fi for Basic Auction Theory\iffinal\thanks{This work has been supported by EPSRC grant EP/J007498/1.  We would like to thank Peter Cramton and Elizabeth Baldwin for sharing their auction designer's point, and Christian Maeder for his recent improvements to Hets.\ifarxiv The final publication is available at \texttt{http://link.springer.com}.\fi}\fi}
\author{Christoph Lange\inst{1}
\and Marco B.\ Caminati\inst{2}
\and Manfred Kerber\inst{1}
\and Till Mossakowski\inst{3}
\and Colin Rowat\inst{4}
\and Makarius Wenzel\inst{5}
\and Wolfgang Windsteiger\inst{6}
\institute{%
Computer Science, University of Birmingham, UK
%\email{math.semantic.web@gmail.com, m.kerber@cs.bham.ac.uk}%
\and \texttt{http://caminati.net.tf}, Italy
%\email{???}%
\and University of Bremen and DFKI GmbH Bremen, Germany
%\email{till.mossakowski@dfki.de}%
\and Economics, University of Birmingham, UK
%\email{c.rowat@bham.ac.uk}%
\and Univ. Paris-Sud, Laboratoire LRI, UMR8623, Orsay, F-91405, France
\and RISC, Johannes Kepler University Linz (JKU), Austria
%\email{wolfgang.windsteiger@risc.jku.at}%
\texttt{http://www.cs.bham.ac.uk/research/projects/formare/code/auction-theory/}
% \texttt{http://www.cs.bham.ac.uk/research/projects/formare/}
}}

\maketitle

\begin{abstract}
  Novel auction schemes are constantly being designed.  Their design has significant consequences for the allocation of goods and the revenues generated.  But how to tell whether a new design has the desired properties, such as efficiency, i.e.\ allocating goods to those bidders who value them most?  We say: by formal, machine-checked proofs.  We investigated the suitability of the Isabelle, Theorema, Mizar, and Hets/\allowbreak CASL/\allowbreak TPTP theorem provers for reproducing a key result of auction theory: Vickrey's 1961 theorem on the properties of second-price auctions.  Based on our formalisation experience, taking an auction designer's perspective, we give recommendations on what system to use for formalising auctions, and outline further steps towards a complete auction theory toolbox.
\end{abstract}

\section{Motivation: Why Formalise Auction Theory?}
\label{sec:motivation}
%\todo{MC@CL: changed (renewbibmacro) biblatex style to display "shortjournal" entry in place of "journal" entry where available. Add "shortjournal" to the relevant .bib records to exploit that (I tested that with one), or comment out renewbibmacro block to revert.\\ CL@MC: Thanks, but sorry, I noticed this too late, and had already shortened journal titles (and a lot of other stuff) manually.}
Auctions are a widely used mechanism for allocating goods and services\footnote{For the US, the National Auctioneers Association reported \$268.5 billion for 2008~\cite{NAA:AuctionsPastPresentFuture10:biblatex}.}, perhaps second in importance only to markets.  They are used to allocate electromagnetic spectrum, airplane landing slots, %bus routes, 
oil fields, bankrupt firms,
works of art, eBay items, and to establish exchange rates, treasury bill yields, and stock exchange opening prices.  Novel auction schemes are constantly being designed, aiming to maximise the auctioneer's revenue, foster competition in subsequent markets, and to efficiently allocate resources.

Auction design can have significant consequences.  Klemperer attributed the low revenues gained in some government auctions of the 3G radio spectrum in 2000 (€20 per capita vs.\ €600 in other countries) to bad design~\cite{kle-04}.  Design practice outstrips theory, especially for complex modern auctions such as combinatorial ones, which accept bids on subsets of items (e.g.\ collections of spectrum%, bus routes, landing slots
).  Designing a revenue-maximising auction is $\mathit{NP}$-complete~\cite{co-sa-04} even with a single bidder.  Important auctions often run “in the wild” with few formal results~\cite{kle-10}.  We aim at convincing auction designers that investing into formalisation pays off with machine-checked proofs and a deeper understanding of the theory.  To this end, we want to provide them with a toolbox of basic auction theory formalisations, on top of which they can formalise and verify their own auction designs – which typically combine standard building blocks, e.g.\ an ascending auction converting to a sealed-bid auction when the number of remaining bidders equals the number of items available.  Given the ubiquity of specialist support across a range of service sectors, we conjecture that auction designers might be supported by formalisation experts, creating a niche for specially trained experts at the interface of the core mechanised reasoning community and auction designers.

Our ForMaRE project (\ul{for}mal \ul{ma}th\-e\-mat\-i\-cal \ul{r}easoning in \ul{e}conomics~%
\cite{LRK:FormareProject13}) seeks to increase confidence in economics' theoretical results, to aid in discovering new results, and to foster interest in formal methods within economics.  To formal methods, we seek to contribute new challenge problems and user experience feedback from new audiences.  Auctions are representative of practically relevant fields of economics that have hardly been formalised so far.\footnote{Even code verification is typically not considered, although % Robert 
Leese, who worked on the UK's spectrum auctions, has called for auction software to be added to the Verified Software Repository at \url{http://vsr.sourceforge.net}~\cite{wo-la-bi-fi-09}.
% best source for claim seems to be http://www.ma.hw.ac.uk/siamstudentchapter/SIAM-Edinburgh-Leese.pdf}
}  Economics has been formalised before~\cite{KLR:EconomistsMechanizedReasoning12}, particularly social choice theory (cf.\ §\ref{sec:related-work} and \cite{ge-en-11}) and game theory (cf.\ \cite{ta-li-11} and our own work~\cite{ke-ro-ww-11}).  However,
none of these formalisations involved economists.  Formalising (mathematical) theories and applying mechanised reasoning tools remain novel to economics.\footnote{There is a field “computational economics”; however, it is mainly concerned with the \emph{numerical} computation of solutions or simulations (cf., e.g., \cite{InitCompEcon}).}

§\ref{sec:requirements} establishes requirements for the Auction Theory Toolbox (ATT); §\ref{sec:approach} explains our approach to building it% from elaborated paper versions of representative problems
.  §\ref{sec:eval} is our main contribution: a qualitative comparison of how well four different theorem provers satisfy our requirements.  §\ref{sec:related-work} reviews related work, and §\ref{sec:conclusion-outlook} concludes and provides an outlook.

\section{Requirements for an Auction Theory Toolbox}
\label{sec:requirements}

Conversations with auction designers established ATT requirements as follows:

\begin{enumerate}[label=\textbf{D\arabic*},ref=D\arabic*]
\item\label{reqD:formalisation} Formalise ready-to-use basic auction concepts, including their definitions and essential properties.
\item\label{reqD:custom} Allow for extension and application to custom-designed auctions without requiring expert knowledge of the underlying mechanised reasoning system.
\end{enumerate}
From a computer scientist's technical perspective, these translate to:
\begin{enumerate}[label=\textbf{C\arabic*},ref=C\arabic*]
\item\label{reqC:lang} Identify the right language to formalise auction theory.  This language should
  \begin{enumerate*}
  \item\label{reqC:lang:expr} be sufficiently expressive for concisely capturing complex concepts, while supporting efficient proofs for the majority of problems,
  \item\label{reqC:lang:learn} be learnable for economists used to mathematical textbook notation, and
  \item\label{reqC:lang:lib} provide libraries of the mathematical foundations underlying auctions.
  \end{enumerate*}
\item\label{reqC:mrs} Identify a mechanised reasoning system
  \begin{enumerate*}
  \item\label{reqC:mrs:devel} that assists with cost-effective development of formalisations,
  \item\label{reqC:mrs:reuse} that facilitates reuse of formalisations already existing in the toolbox,
  \item\label{reqC:mrs:understand} that creates comprehensible output to help users understand, e.g., why a proof attempt failed, or what knowledge was used in proving a goal, and
  \item\label{reqC:mrs:community} whose community is supportive towards users with little specific technical and theoretical background.
\end{enumerate*}
\end{enumerate}
Note the conflicts of interest: a single language might not meet requirement~\ref{reqC:lang:expr}, and if it did, it might not be supported by a user-friendly system.

\section{Approach to Building the Auction Theory Toolbox}
\label{sec:approach}

To avoid a chicken-and-egg problem, we identify relevant domain problems in parallel to identifying languages and systems suitable for formalisation.

\subsection{The Domain Problem: Vickrey's Theorem and Beyond}
\label{sec:nail}

We started with %William
Vickrey's 1961 theorem on the properties of second-price auctions of a single, indivisible good, whose bidders' private values are not publicly known.  Each participant submits a sealed bid; one of the highest bidders wins, and pays the highest \emph{remaining} bid; the losers pay nothing.  Vickrey %, who was awarded economics' Nobel prize in 1996 for his work,
proved that “truth-telling” – submitting a bid equal to one's actual valuation of the good – was a \emph{weakly dominant} strategy, i.e.\ that no bidder can do strictly better by bidding above or below their valuation \emph{whatever} the other bidders do.  Thus, the auction is also \emph{efficient}, allocating the item to the bidder with the highest valuation.  Bidders only have to know their own valuations; in particular they need no information about others' valuations or the distributions these are drawn from.

As variants of Vickrey auctions are widely used (e.g.\ by eBay, Google and Yahoo!~\cite{wikipedia:VickreyAuction}), this formalisation will enable us to prove properties of contemporary auctions as well.  The underlying theory is straightforward to understand even for non-economists and can be formalised with reasonable effort.  Finally, formalising Vickrey provides a good introduction for domain experts to mechanised reasoning technology by serving as a small, self-contained showcase of a widely known result%, arousing their interest in further possibilities
%.  Thus, re-establishing it may
, helping to build trust in this new technology.

%Eric
Maskin collected %high level versions of 
13 theorems, including Vickrey's, in a review~\cite{mas-04} of an influential auction theory textbook~\cite{mil-04}.  This sets the roadmap for building the ATT – a collaborative effort, to which we welcome community contributions\todo{CR: more explicit statement how? E.g.\ wiki-style? Concrete invitations are easier to accept than are open-ended ones.\\ CL@CR: right in theory, but IMHO irrelevant in practice (and we can't even afford one more \emph{word} here). The cost/benefit ratio of seriously maintaining Planetary or any other wiki is low, \cite{AuctionTheoryToolbox} has clear instructions how to get in touch with us, and the best (proven!) way to attract contributions is the one we pursued with this paper: personal invitation with clear benefits for the contributors.}~\cite{AuctionTheoryToolbox}.

\subsection{Paper Elaboration to Prepare the Machine Formalisation}
\label{sec:prep-paper-form}

To prepare the machine formalisation, we refined the original paper source, aware that current mechanised reasoning systems typically require much more explicit statements than commonly found on paper: automated provers must find proofs without running out of search space, whereas proof checkers require proofs at a certain level of detail, which in turn requires detailed statements.  Maskin states Vickrey's theorem in two sentences and proves it in another six sentences~\cite[Proposition~1]{mas-04}.\footnote{%Note that 
The high level of Maskin's text is owed to its summative nature.  Original proofs in auction theory are typically more thorough.}  Our elaboration uses eight definitions specific to the domain problem plus an auxiliary one about maximum components of vectors, as follows:

{\termdef{participant}{participant}
\termdef{seller}{seller}\termdef{allocation}{allocation}
\termdef{payment}{payment}
\symdef{alloc}[1]{\component{x}{#1}}
\symdef{component}[2]{{#1}_{#2}}
\symdef{devVec}[3]{{#1}^{{#2}{\leftarrow}{#3}}}
\symdef{pay}[1]{\component{p}{#1}}
\symdef{maximum}[1]{\overline{#1}}
\symdef{maximumExcept}[2]{\overline{#1}_{-#2}}
\symdef{val}[1]{\component{v}{#1}}
$N = \left\{ 1, \ldots, n \right\}$ is a set of \emph{{\participant}s}, often indexed by $i$. An \emph{\allocation} is a vector $\bm{x} \in \left\{ 0, 1 \right\}^n$ where $\alloc{i} = 1$ denotes \participant $i$'s award of the indivisible good to be auctioned (i.e.\ “$i$ wins”), and $\alloc{j} = 0$ otherwise.
An \emph{outcome} $\left( \bm{x}, \bm{p} \right)$ specifies an \allocation and a vector of {\payment}s, $\bm{p} \in \mathbb{R}^n$, made by each \participant $i$.
\capitalize\participant $i$'s \emph{payoff} is $u_i \equiv \val{i} \alloc{i} - \pay{i}$, where $\val{i} \in \mathbb{R}_+$ is $i$'s valuation of the good.
A \emph{strategy profile} is a vector $\bm{b}\in\mathbb{R}^n$, where $\component{b}{i} \ge 0$ is called $i$'s \emph{bid}.\footnote{This simplification is sufficient for proving the theorem.  More precisely, all participants know that each $\val{i}$ is an independent realisation of a random variable with distribution density $f$.  A \participant's \emph{strategy} is a mapping $g_i$ such that $\component{b}{i} = g_i \left( \val{i}, f \right)$.}
\todo{CL: either devise a more readable notation here, or don't use overbar in combination with hat.  Or reduce proof to structural sketch.}For an $n$-vector $\bm{y} = \left( \component{y}{1}, \ldots, \component{y}{n} \right) \in \mathbb{R}^n$, let
       $\maximum{y} \equiv \max_{j \in N} \component{y}{j}$ and $\maximumExcept{y}{i} \equiv \max_{j \in N \backslash \left\{ i \right\}} \component{y}{j}$.

\begin{definition}[Second-Price Auction]\label{def:spa}
Given $M \equiv \left\{ i \in N : \component{b}{i} = \maximum{b} \right\}$, a \emph{second-price auction} is an outcome $\left( \bm{x}, \bm{p} \right)$ satisfying:
     \begin{enumerate}[nosep]
     \item $\forall j \in N \backslash M, \alloc{j} = \component{p}{j} = 0$; and
     \item for one\footnote{When running an auction in practice, this $i$ may be selected randomly%according to any randomisation device
 , but this circumstance does not matter for the proof of Vickrey's theorem.} $i \in M$, $\alloc{i} = 1$ and $\pay{i} = \maximumExcept{b}{i}$, while, $\forall j \in M \backslash \left\{ i \right\}, \alloc{j} = \component{p}{j} = 0$.
     \end{enumerate}
   \end{definition}
\begin{definition}[Efficiency]\label{def:efficiency}
An efficient auction maximises $\sum_{i \in N} \val{i} \alloc{i}$ for a given $v$, i.e., for a single good, $\alloc{i} = 1 \Rightarrow \val{i} = \maximum{v}$.\todo{CL@CR: MK reformulated this from the original $v_i \ne \bar{v} \Rightarrow x_i = 0$. Isn't it more intuitive this way? CR@CL: we may want to encode at some point the idea that only one $i$ can have $x_i = 1$\\ CL: Def.~\ref{def:spa} implies this (but rather implicitly).  BTW we could have used $\exists!$ here – most of our formalised languages have this built in as well.}
\end{definition}

\begin{module}
  \begin{definition}[Weakly Dominant Strategy]\label{def:WDS}
    Given some auction, a strategy profile $\bm{b}$ supports an \emph{equilibrium in weakly dominant strategies} if, for each $i \in N$ and any \todo{CL: hat looks ugly with overbar (as used in the proof), so consider an alternative to hat}$\hat{\bm{b}} \in \mathbb{R}^n$ with $\component{\hat{b}}{i} \ne \component{b}{i}$,
      $u_i \left( \component{\hat{b}}{1}, \ldots, \component{\hat{b}}{{i-1}}, \component{b}{i}, \component{\hat{b}}{i+1}, \ldots, \component{\hat{b}}{n} \right) \ge u_i \left( \hat{\bm{b}} \right)$.\footnote{The notation $u_i \left( \bm{b} \right)$ is standard in economics but formally misleading.  A more careful notation is $u_i \left( \alloc{i}, \val{i}, \pay{i} \right)$, where $\alloc{i}$ and $\pay{i}$ depend on $\bm{b}$ and the auction type.}
    I.e., whatever others do, $i$ will not be better off by deviating from the original bid $\component{b}{i}$.
  \end{definition}
\end{module}

\begin{module}[id=vickrey]
  \importmodule{second-price-auction}
  \importmodule{auction-properties}
  \begin{theorem}[Vickrey 1961; Milgrom 2.1]\label{RefVick}
    In a second-price auction, the strategy profile $\bm{b} = \bm{v}$ supports an equilibrium in weakly dominant strategies.  Furthermore, the auction is efficient.
  \end{theorem}

The attempt to be close to a paper formalisation may introduce artefacts that unnecessarily complicate machine formalisation.  E.g., the contiguous numeric participant indexing is merely a convention: formally any relation between participants' valuation, bid, allocation, and payment vectors suffices.  Similarly, the product $\val{i} \alloc{i}$ recalls the general divisible good case ($\alloc{i}\in [0,1]$) and works around the lack of an easy and compact “if–then–else” textbook notation.\footnote{Case distinctions with curly braces consume at least two lines.%  If $w$ were the winner's index, one could use $\delta_{i,w}:=\begin{cases}1 & i=w\\ 0 & \mathrm{otherwise}\end{cases}$, which is less widely understood.
}

  \begin{proof}\label{RefProof}
    Suppose participant $i$ bids $\component{b}{i} = \val{i}$, whatever $\hat{b}_j$ the others bid.  Let $\devVec{\hat{\bm{b}}}{i}{v}$ abbreviate the overall vector $%\left
(\hat{b}_1, \ldots, \hat{b}_{i-1}, \component{v}{i}, \hat{b}_{i+1}, \ldots, \hat{b}_n%\right
)$.  \todo{CL: maybe only expose the \emph{structure} of the proof; refer to ATT homepage for full detail}There are two cases\footnote{Our initial elaboration of Maskin's proof, which distinguishes cases on the basis of participants' bids, resulted in nine leaf cases.  Straightforward on paper, we found them tedious to formalise in Isabelle, which triggered the rearrangement shown here.}:
    \begin{enumerate}[nosep]
    \item $i$ wins.  This implies  $\component{b}{i}  = \val{i} = \maximum{\devVec{\hat{\bm{b}}}{i}{v}}$, $\pay{i} = \maximumExcept{\devVec{\hat{\bm{b}}}{i}{v}}{i}$,
      and $u_i(\devVec{\hat{\bm{b}}}{i}{v}) = \val{i} - \pay{i} = \component{\devVec{\hat{\bm{b}}}{i}{v}}{i} - \maximumExcept{\devVec{\hat{\bm{b}}}{i}{v}}{i} \ge 0$.  Now consider $i$ submitting an arbitrary bid $\component{\hat{b}}{i} \ne \component{b}{i}$, i.e.\ assume an overall bid vector $\hat{\bm{b}}$.  This has two sub-cases:
      \begin{enumerate}[nosep]
      \item $i$ wins with the other bid, i.e.\ $u_i(\hat{\bm{b}}) = u_i(\devVec{\hat{\bm{b}}}{i}{v})$,
        as the second highest bid has not changed.
      \item $i$ loses with the other bid, i.e.\ $u_i(\hat{\bm{b}}) = 0 \le u_i(\devVec{\hat{\bm{b}}}{i}{v})$.
      \end{enumerate}
    \item $i$ loses. This implies $\pay{i} = 0$, $u_i(\devVec{\hat{\bm{b}}}{i}{v}) = 0$, and
      $\component{b}{i} \le \maximumExcept{\devVec{\hat{\bm{b}}}{i}{v}}{i}$; otherwise $i$ would have won. This yields
      again two cases for $i$'s alternative bid :
      \begin{enumerate}[nosep]
      \item $i$ wins, i.e.\
        $u_i(\hat{\bm{b}}) = \val{i} - \maximumExcept{\hat{\bm{b}}}{i} = \component{b}{i} - \maximumExcept{\devVec{\hat{\bm{b}}}{i}{v}}{i}  \le 0 = u_i(\devVec{\hat{\bm{b}}}{i}{v})$.
      \item $i$ loses, i.e.\ $u_i(\hat{\bm{b}}) = 0 = u_i(\devVec{\hat{\bm{b}}}{i}{v})$.
      \end{enumerate}
    \end{enumerate}
    By analogy for all $i$, $\bm{b} = \bm{v}$ supports an equilibrium in weakly dominant strategies.  Efficiency is immediate: the highest bidder has the highest valuation.\qed
  \end{proof}
\end{module}}

\subsection{Choosing Language and System}
\label{sec:choos-mech-reas}

In terms of \emph{logic}, it is not immediately obvious whether Vickrey's theorem is inherently higher-order.  Defining the maximum operator on arbitrarily sized finite sets of real-valued bids and proving its essential properties requires induction and thus exceeds first-order logic (FOL): similarly for the finiteness of a set\footnote{Finiteness matters: the set $\{b_i=1-\frac{1}{i} : i=1,2,3,\dots\}$ has no maximum.} and a formalisation of real numbers.\footnote{Real numbers are not usually required for running auctions in \emph{practice}.  Even financial exchanges that allow “sub-pennying” have a minimal discrete quantum of currency.}
% \todo{MC: Added this period.\\ CL: OK to have a concrete example – but could you please replace, or accompany, those technical names (LM12, NAT\_1:39, NAT\_1:sch 2) by some descriptive text that one can understand without looking into the sources?}
However, if one takes real vectors and a maximum operation on them for granted, and explicitly requires the maximum to exist, FOL suffices to formalise the relevant domain concepts: single good auctions, second-price auctions, and the theorem statement.\todo{CL@MC: shortened this; is it still correct?  And omitted the listing.}\footnote{For instance, our \M{} proof never invokes any second-order \emph{scheme}
% (see footnote \ref{RefScheme} on page \pageref{RefScheme})
directly.
Two proof steps use the fact that a finite set of numbers includes its maximum, which is proved in the Mizar Mathematical Library (\mml{}) using the induction scheme%\lstinline[language=Mizar]|for k being Nat holds P[k] provided P[0]|\allowbreak\lstinline[language=Mizar]|and (for k be Nat st P[k] holds P[k+1]);|
.}

In terms of \emph{syntax}, we assume that auction designers will prefer a language that is close to the textbook mathematics they are used to, \todo{CL@All: FYI Elizabeth Baldwin actually said that if it is close to Matlab or C this would be an advantage.  But OTOH I argued that we are planning to offer a substitute for paper proofs in the first place, not yet for algorithm implementations.}rather than having a programming language flavour.  We assume that at least optional type annotations support intuitive modelling of domain concepts (e.g.\ an auction as a function that takes bids and returns an allocation and payments) and prevent formalisation mistakes by cheap early checks (cf.~\cite{LamportPaulson:SpecLangTyped99}).

In terms of \emph{user experience}, we study two paradigms: \emph{automated provers} try, given a theorem and a knowledge base, to automatically find a proof, potentially appealing to our audience if the user just has to push a button (as with model checkers).  \emph{Interactive provers} interactively check a proof written by the user, which may be convenient when a paper proof already exists.

\section{Qualitative Comparison of the Languages and Systems}
\label{sec:eval}

We have formalised Vickrey's theorem in four systems% so far
, which differ in logic, syntax and user experience:  %% FOR AN ECONOMICS PUBLICATION: \todo{CL: maybe mention somewhere that Mizar form.\ has more general definition of bids; efficiency for infinite participants with multiple bids (@CR: makes sense? CR@CL: not inf. participants WITH multiple bids – each? – CL@MC: please clarify!); equilibrium for infinite participants. MC@CL/CR: Currently, equilibrium is formalised wrt an arbitrary number of participants, but each placing one bid; moreover, the bids as a whole have to admit an upper bound. I think the first requirement is admissible, and should have that ready soon.\\ CL: sounds good to me.  OTOH in \emph{this} paper (given its prover comparison focus) I don't find a good place to mention this.  We should mention this in an economics publication. MC@CL: Yes, sounds reasonable to me.}
Isabelle%\footnote{There was no specific conceptual reason for starting with Isabelle.}
, followed by \M{}, CASL and Theorema.  For each system at least one author has in-depth knowledge.  The purpose of redoing formalisations from scratch is to understand the specific advantages and disadvantages of the systems and to obtain as idiomatic a formalisation as possible.  %\footnote{Hets is, for example, capable of translating CASL to Isabelle, but the resulting Isabelle code would not make use of HOL features other than inductive datatypes.}
The formalisations and instructions for using them are available from the ATT homepage~\cite{AuctionTheoryToolbox}% under a free dual licence for source code (ISC, %%a cleaned-up BSD licence
% similar to BSD) and creative works (CC-BY), as inspired by the considerations for licencing the \mml~\cite{AlaKohMam:lmml11}.
.  \todo{CL: FYI I know this table has lots of footnotes (cut some already!), but we would get new formatting problems by removing it from its own landscape page.  And space still doesn't permit integrating the other table into that same landscape page.}Tab.~\ref{tab:systems} compares the features of the systems and their languages and shows the \todo{CL@TM: please update CASL/TPTP state once you made progress. TM: done}state of our work.  The following subsections assess the languages and systems w.r.t.\ the technical requirements C* of §\ref{sec:requirements}.  Tab.~\ref{tab:comparison} at the end of this section summarises our findings to underpin our final recommendations.

\begin{sidewaystable}
\def\extraverticalspace{\addlinespace[.5ex]}
\caption{Languages and systems we compared; state of our formalisations}
\label{tab:systems}
\centering
\begin{varwidth}{\linewidth}
\renewcommand*\footnoterule{}
\begin{tabularx}{\linewidth}{
>{\hsize=.65\hsize\RaggedRight}X% language
>{\hsize=.8\hsize\RaggedRight}X% logic
>{\hsize=.35\hsize\RaggedRight}X% prover
>{\hsize=2.7\hsize\RaggedRight}X% UI
>{\hsize=.75\hsize\RaggedRight}X% licence
>{\hsize=.75\hsize\RaggedRight}X% formalisation progress
}
  \FL
  Language & Logic & Prover & User Interface & Licence
  & Formalisation
  \ML
Isabelle/HOL\newline 2013~\cite{isabelle} & HOL (simply-typed set theory) & \hspace{0pt}interactive\footnote{Isabelle integrates internal and external automated provers.} & document-oriented %Prover
IDE (Isabelle/jEdit~\cite{Wenzel:IsabelleJEdit12}) or programmer's text editor (Proof General Emacs~\cite{DA:PG}) & BSD/LGPL/\allowbreak GPL & complete incl.\ proof
  \NN
\extraverticalspace Theorema\newline 2.0~\cite{Windsteiger:Theorema20UI12} & FOL\newline + set theory\footnote{Theorema actually supports HOL.  We, however, just needed FOL besides the built-in sets, tuples, and the $\max$ operator.} & \hspace{0pt}automated\footnote{For each goal, the prover can be configured individually.} & textbook-style documents, proof management GUI (add-on for Mathematica CAS) & GPL%v3
\footnote{Theorema is under GPL but needs the commercial, closed-source Mathematica.  Economists tend to be pragmatic about that.}
  & statements com-\newline plete, no proof\footnote{Theorema is in transition to the new 2.0.  Its architecture, inference engine, and user interface are fully implemented, but its collection of \emph{inference rules} is still incomplete. Therefore, the proof does not yet work.}
  \NN
\extraverticalspace\M\newline 8.1.01~\cite{grabowski2010mizar} & FOL\footnote{\emph{Schemes}\label{RefScheme} permit a limited degree of higher-order reasoning%; however, those are so constrained that \M\ has been termed a “$1.001$th-order prover”~\cite{wiedijk2006writing}
.}\newline + set theory & batch verifier & CLI\footnote{The \emph{verifier} produces a list of numerical errors codes and their source file positions.
  The ancillary utilities \emph{errflag} and \emph{addfmsg} decorate source files with this information, and optionally append terse textual explanations of the relevant error codes.}; programmer's text editor (Emacs add-on) & freeware/GPL%v3
+\allowbreak CC-BY-SA\footnote{The \M\ proof checker is closed-source; the MML is free%~\cite{AlaKohMam:lmml11}
.}
  & complete incl.\ proof
  \NN
\extraverticalspace CASL/\newline TPTP\footnote{\label{fn:CASL}Common Algebraic Specification Language.  “CASL/TPTP” denotes our use of CASL as an input language for automated FOL provers (here: SPASS, E, Darwin) using the TPTP~\cite{Sutcliffe:TPTP09} exchange language.  CASL features some second-order features, e.g.\ inductive datatypes.}~\cite{CASL} & sorted FOL\textsuperscript{\ref{fn:CASL}} & auto\-mated\footnote{The proof is largely automatic. However, Vickrey's theorem is too complex to for automated proving in one step. Thus, the proof script introduces auxiliary lemmas and selects suitable axioms and provers for proving them. Proof times range from fractions of seconds if the exact list of axioms used is known beforehand to hours if not. However, once a proof is found, the prover can output the list of axioms used and thus speed up subsequent replays of the proof.} & progr.'s text editor (Emacs add-on), proof mgmt.\ GUI+\allowbreak CLI (Hets 0.98\footnote{Heterogeneous tool set; gives access to a wide range of automated theorem provers.  We use FOL provers, most of which share the unsorted TPTP FOF~\cite{Sutcliffe:TPTP09} as a common input format.  Hets translates CASL to FOF by introducing auxiliary predicates for sorts.}~\cite{HETS:on}), web service (System on TPTP~\cite{SystemOnTPTP}) & GPL%v2
  & complete incl.\ proof
\LL
\end{tabularx}
\end{varwidth}
\end{sidewaystable}

\subsection{Level of Detail and Explicitness Required (req.\ \ref{reqC:lang:expr})}
\label{sec:level-of-detail}

All systems required greater detail and explicitness than the paper elaboration of §\ref{sec:prep-paper-form}.  The Isabelle formalisation needs 3 additional definitions %
% SecondPriceAuction:
% * DELETED second_price_auction_winners_payment
% * second_price_auction_winner
% * second_price_auction_loser
% SingleGoodAuction:
% * payoff_vector
and 7 auxiliary lemmas.
% SecondPriceAuction:
% * allocated_implies_spa_winner
% * not_allocated_implies_spa_loser
% * only_max_bidder_wins
% * second_price_auction_winner_payoff
% * second_price_auction_loser_payoff
% * winners_payoff_on_deviation_from_valuation
% SingleGoodAuction:
% * valuation_is_bid
\todo{CL: TODO update once done with the Theorema and CASL formalisations\\ @TM: maybe you can do it for CASL once you have new results}Guiding the automated provers of Theorema and Hets and \M's proof checker required similar numbers of auxiliary statements, plus, in Theorema and Hets, further ones to emulate proof steps (cf.\ §\ref{sec:expr-vs.-effic}).
However, first steps beyond Vickrey's theorem suggests that these auxiliaries make it easier to formalise \emph{further} notions.  As our work involved beginners and experts\footnote{The \M\ formalisation was, e.g., completely written by an expert (Caminati), whereas the Isabelle formalisation was initially written by a first-time user with a general logic background (Lange), then largely rewritten by an expert (Wenzel).}, we can only approximately quantify the formalisation effort beyond the paper elaboration.  The “de Bruijn factor”~\cite{Wiedijk:tdbf:on}, the formalisation size divided by the size of an informal {\TeX} source, measured after stripping comments and \textit{xz} compression,
is around $1.5$ for all formalisations\footnote{A typical average is $4$, but our paper proof is particularly detailed.} except Theorema\todo{CL@WW: please check whether you like this.}\footnote{Determining a de Bruijn factor for Theorema does not make sense: single keystrokes or clicks may yield complex inputs, Mathematica notebooks store layout and maintenance information%, which cannot be removed fully
, and Theorema caches proofs in the notebook (cf.\ §\ref{sec:output}).}.  This observation suggests that machine formalisation is generally still harder than elaboration on paper.

\todo{MC: Adapted from AISB paper; feel free to take that away.\\ CL@MC: I merged this with some of the text that we had after the paper formalisation; please check if I still got your message right.}Even while explicit machine formalisation imposes tedious work on the author, it can also prove beneficial.  On paper, it was neither immediately obvious that exactly one participant wins a second-price auction\todo{CL@All: not sure whether to say anything else about this: most languages have a convenient “exists exactly one” built-in.}, nor that the outcome is a function of the bids.  While obvious that at least two participants are required to define the “second highest bid”, the standard literature largely overlooks this, but formalisation forced us to choose whether to allow it (by, e.g., defining $\max \emptyset\equiv 0$) or to explicitly require $n\ge 2$.%  CUT: POINT IS MADE; EXAMPLE FEELS CONTRIVED

\subsection{Expressiveness vs.\ Efficiency (req.\ \ref{reqC:lang:expr})}
\label{sec:expr-vs.-effic}

As discussed in §\ref{sec:prep-paper-form}, we did not strictly take the elaborated paper source as a specification for the formalisation, but wrote idiomatic formalisations.  In Isabelle and \M{}, we, e.g., avoided specific intervals $\{1,\dots,n\}$ as sets of auction participants: arbitrary (finite) sets of natural numbers simplify the formalisation, and the highest and second highest bids are determined using library set operations.  In contrast, Theorema naturally indexes its built-in tuples from $1$ to $n$ and allows for restricting quantified variables to such ranges, e.g.\ $\forall_{i=1,\dots,n}$.

The CASL formalisation confirms the assumption of §\ref{sec:choos-mech-reas} that FOL suffices for expressing and proving the essence of Vickrey's theorem.  For many FOL provers, CASL's (sub)sorts\footnote{TPTP's typed first-order form (TFF~\cite{SSCB:TPTP-FOF12}) is sorted, but without subsorts.  We have not used it, as Hets cannot currently produce it from CASL.} \todo{CL: removed “and inductive datatypes” for now: these really go beyond FOL.  But, @TM, I'd like to somehow mention that we can still \emph{use} them (as we discussed by email and Skype).}are mere syntactic sugar but allow us to stay close to the domain language, speaking, e.g., of “valuation vectors”, each of which also is a valid “bid vector”.  Note that we have avoided using partial functions (e.g., for modelling out-of-scope vector indices) because of the complex logic translations required for coding them out.

Isabelle and Mizar process the proof in a few seconds on a 2.5 GHz dual-core processor; Hets/TPTP need about an hour%
% MC: $10$ seconds on a Celeron @ 1.5 GHz.
\todo{CL: TODO in the end fill in an actual low common upper bound for all systems. TM: done}; in Theorema it is \todo{CL@TM: please update if you manage to complete it. TM: done}not yet complete.  We used rather weak HOL features, e.g., no synthesisation of functions.  Coinciding with earlier, general observations on HOL~\cite{Farmer:STT08}, the low processing time suggests that there is no disadvantage in choosing a rich logic, which allows for expressing relevant concepts (such as maxima of finite sets of real numbers) naturally.  Our formalisations' small size (less than 5 K after compression) does not yet warrant a precise quantitative judgement of time efficiency.  Particularly for FOL there exist highly optimised automated provers.  They are conveniently accessible in Hets, via the System on TPTP~\cite{SystemOnTPTP} web service (accepting TPTP input that Hets can generate), but also from Isabelle/HOL via the Sledgehammer interface (see §\ref{sec:proof-management}).  Still, we observed a source of inefficiency in formalising for automated provers: the high share of preconditions with long conjunctions in our CASL formalisation makes it hard for the automated FOL provers to identify applicable axioms.  Such conjunctions result from the absence of structured proofs in CASL.  This requires, whenever a theorem is too complex for automated proving, to “emulate” proofs steps via auxiliary lemmas, whose antecedents are conjunctions of all relevant assumptions in the current branch of the proof tree.  Performance improvements by guiding provers through the search space can, however, be achieved with the extra effort of grouping frequently occurring conjunctions of assumptions into single abstract predicates, as in the following concrete case for the proof of Vickrey's theorem: $\mathit{spaWithTruthfulOrOtherBid}(n, x, p, v, \hat{\bm{b}}, i, \bm{b})\Leftrightarrow\mathit{secondPriceAuction}(n, x, p)\wedge|v|=|\hat{\bm{b}}|=n\wedge\mathit{inRange}(n, i)\wedge\hat{\bm{b}}_i\ne v_i\wedge \bm{b}=\hat{\bm{b}}[i{\leftarrow}v]$.

\subsection{Proof Development and Management (req.\ \ref{reqC:mrs:devel})}
\label{sec:proof-management}

% \todo{MC: Should we mention the difference btw declarative and procedural systems?
% That aspect impacts workflow and readability to a large extent, to me.
% I heard that some systems have utilities to `translate' from procedural proof to a declarative one, but IIUC this only affects output, not workflow.}
The systems we studied offer different ways of invoking automated provers and keeping track of proof efforts in progress.
The “apparent” difference between automated and interactive theorem proving blurs at a closer look.  The interactive prover Isabelle features various automated proof methods% as part of the standard HOL library
; furthermore Sledgehammer gives access to E, SPASS, and TPTP provers.  One can configure the facts they should take into account (e.g.\ local assumptions and conclusions).  \todo{CL@MC: moved this sentence here: OK?}For \M{}, there are also automated external tools (MPTP, MoMM, Miz$\mathbb{A}\mathbb{R}$)~\cite{rudnicki2011escape}.
Theorema's automated proving workflow is conceptually similar: specifying the knowledge to be used, then configuring the prover.\todo{CL@WW: I had to shorten your text.  Here, we are mainly comparing Theorema 2.0 to other systems; therefore improvements of 2.0 over 1 are not the most central thing to be mentioned.  What you said about generating human-comprehensible proofs IMHO rather belongs into the “output” section; I merged it into what we already had there.}\footnote{For Theorema, a prover is a \emph{collection of inference rules} applied in a certain \emph{strategy}.}  %\footnote{The easy selection and fine-tuning of the rules and the strategy are among the big improvements in Theorema 2.0 compared to Theorema 1.}
%, with the aim to automatically generate a \emph{human-comprehensible proof} that appears like \emph{written by a well-trained mathematician}.
Hets users can select axioms and previously proved theorems to be sent to an automated prover but have little control beyond.  Isabelle's prover configuration is editable within the formalisation source.  Theorema stores it in hidden fields within the formalisation and exposes it via a dedicated GUI.  Configuring proof tools in Hets is separate from the formalisation: the proof management GUI does not currently store settings persistently; however one can write scripts to be processed on the command line.

Just as Isabelle requires complex statements to be proved in multiple steps, involving different proof methods, the automated provers of Theorema\footnote{This assessment relies on experience with Theorema 1.} and Hets also require guidance by explicit configuration at times, as can be seen from the \texttt{*.hpf} proof scripts in our Hets formalisation~\cite{AuctionTheoryToolbox}.  Often, a theorem $c: A\Rightarrow C$ was too complex for automated proving, whereas the job could be done by a script that first proved auxiliary lemmas $a: A\Rightarrow B$ and $b: B\Rightarrow C$, possibly with different provers, and then proved $c$ providing only $a$ and $b$ as axioms.  This is conceptually the same as in Isabelle but has four significant user experience differences:
\begin{enumerate*}
\item Each additional “proof step” has to be stated as a lemma with full assumptions on the left hand side (similar to the example in §\ref{sec:expr-vs.-effic}),
\item CASL, originally a specification rather than a prover language, does not syntactically distinguish theorems from lemmas,
\item the scripts have to be maintained separately from the formalisation, and
\item a multi-step proof takes many seconds longer, as Hets translates the input theory from CASL to the respective prover's native language before each proof.\footnote{This is necessary as, by default, each successful proof adds one theorem to the theory.}
\end{enumerate*}
This gives a clear incentive to eliminate unnecessary proof steps from a CASL formalisation.  This experience also influenced our Isabelle formalisation, where writing multi-step proofs is comparatively painless.  There, one lemma had a three-step proof, until experiments with the CASL formalisation made us attempt an automated proof\todo{CL: FYI I said “attempt” because actually I couldn't reproduce this with Hets any more.  The only trace is winners\_payoff\_on\_deviation\_from\_valuation in Isabelle r373.}.  Thus we realised that we could reduce the Isabelle proof to a single step.\footnote{As it makes use of one definition and two lemmas, this was not obvious a priori.}

Mizar differs by focusing, instead of built-in tactics and automated proof methods, on a natural deduction style which
“tries to “keep a low profile” in its logical foundations” and aims at “clarity, human readability and closeness to standard mathematical proofs”~\cite{Urban:MizarMode}.  Influenced by Mizar, the Isar language (“intelligible semi-automated reasoning”) replaced Isabelle's original tactic interface.
In the name of its readability focus, \M{} deliberately prevents users from \todo{CL@MC: cut word “arbitrarily”; hope it's still correct}%arbitrarily 
extending the verifier's power \cite[\S2.1]{Urban:MizarMode}, often
forcing them to justify trivial passages.
\M{}'s \emph{registrations} do allow for custom automation~\cite{CaminatiJar2011}; however, these at times involute exploits often push registrations beyond their intended scope~\cite{kornilowicz2012rewriting} and \todo{CL@MC: shortened the rest of this sentence; hope it's still correct}may result in implicit inferences and less readable proofs.%, and come at a price: inferences may become implicit% in registrations
%, with proofs likely resulting less readable.

Particularly in developing the proof of a theorem as complex as Vickrey's top-down, it is useful to defer proofs of lemmas or proof steps, as to use them in a larger proof without the workaround of temporarily declaring them as axioms.  Theorema proofs can use unproved theorems as knowledge%
% CL@WW: Does it not even have an easy way of making sure that everything is proved?  If so, that would be bad. WW: we don't have this at the moment. It is not in my focus of interest, but it would not be a difficult thing, since I know the proof status for every formula in the system. So I would just need to traverse through all formulae needed in the main proof. I would not mention it in the paper.
.  Isabelle's \lstinline|sorry| keyword creates a fake proof.  CASL theorems are formulas with the annotation \lstinline|%implied|.
When imported into a theory, (open) theorems become axioms, and Hets can use them without proof,
but the open proof obligation is still visible in the imported theory. 
\M's verifier offers top-down proving for free by marking unaccepted inferences
% the verifier marks any statement it is not able to deduce from the given assumptions
as errors %with the specific error code $4$ 
\emph{and then proceeding}% with the remaining statements
.
This results in a formal proof \emph{sketch}, “very close to informal mathematical English” but still close to a fully formalised proof~\cite{Wiedijk04formalproof}.
Furthermore%, if desired%
% (e.g., because the verifier spends too much time trying to “understand” some not yet proved statement)
, one can prefix the keyword \lstinline|proof| with \lstinline|@| to expressly and silently skip a proof, or disable the verifier on arbitrary code portions using pragmas.
Mizar's Emacs mode exposes these % explicit skipping mechanisms
as one-touch macros, which speeds up the verification process and improves interaction~\cite{Urban:MizarMode}.

\subsection{Library Coverage and Searchability (reqs.\ \ref{reqC:lang:lib}, \ref{reqC:mrs:reuse})}
\label{sec:library-coverage}

To a varying degree we have been able to reuse mathematical foundations from the systems' libraries.  %\todo{CL@MW: Actually these Isabelle tools also query (in contrast to Mizar) the \emph{user}'s code, but as we are focusing on the \emph{library} here, I'll skip this to save space.}
Isabelle can \emph{find} reusable material by \lstinline|find_theorems| queries; Sledgehammer helps to extract a sufficient set of lemmas from the library, which is then minimised towards a necessary set.  \todo{CL@MC: Did I get this right?}MML Query is a %comprehensive
search engine for the MML~\cite{Bancerek:irrmq06}.  CASL's library is searchable as plain text; Theorema's is not.

Theorema has a built-in tuple type, including a maximum operation, we used it to formalise bid vectors.  The CASL library provides inductive datatypes such as arrays~\cite{CASL-RM} %\todo{CL: We are not currently \emph{using} them though; not yet sure whether I should elaborate on this, but in any case pure FOL can't deal with them, so our choice to use inductive datatypes (but not the full axiomatisations from the library) is mainly for notational convenience.  @TM: Any further views on this? TM: If CASL is mainly used for TPTP, this is OK. Otherwise, inductive proofs should be mentioned.\\ CL: OK, leaving it at “syntactic sugar” for now, as we are really only using CASL for TPTP, and not doing any induction proofs, e.g., in Isabelle-translated-from-CASL.}
but no $n$-argument maximum operation.  The Isabelle/HOL library provides a \emph{Max} operation on finite sets, and various Cartesian product types suitable for representing bids.  Given Isabelle's functional programming syntax we found it, however, most intuitive to model our own vectors as functions $\mathbb{N}\to\mathbb{R}$ evaluated \todo{CL@All: I cut “evaluated \emph{for arguments} up to a given $n$” – is it still comprehensible this way?}up to a given $n$.  Wrappers make the set maximum operator work on these vectors and prove the properties required subsequently.
Our \M{} formalisation draws on generic relations and functions, which the MML richly covers.  Thus, we only had to add a few interfacing lemmas%
% to interface the existing results to our particular needs
.

\subsection{Term Input Syntax (req.\ \ref{reqC:lang:learn})}
\label{sec:input-syntax}

\todo{CL@MC: As this section focuses on \emph{terms}, you may want to merge some of your text on Mizar's proof language (which I commented in the source) into \ref{sec:proof-management}, but please try without making the text longer.}Conversations with auction designers suggest that they find Theorema's term input syntax most accessible.  The two-dimensional notation in Mathematica notebooks is similar to textbook notation% except for its affinity to square brackets
, and our target audience is largely familiar with Mathematica.  The syntax of Isabelle and CASL is closer to programming languages.  Isabelle's functional type syntax $f:A\Rightarrow B\Rightarrow C$ looks less closely related to textbook notation than CASL's $f:A\ast B\to C$.  %\todo{CL@MC: I hardly dare asking this question (as it will mean adding another sentence), but I'm sure Mizar supports something like this as well. MC@CL: Are you referring to `mixfix' definitions? If so, yes. \M{} also supports assigning arbitrary associative priorities, which I never exploited (IIRC Isabelle does, too), and synonyms-antonyms.\\ CL: Thanks, just described it in very few words for now.  Associative priorities, if we mean the same, exist in CASL and Isabelle.  Synonyms/antonyms: Interesting; Isabelle and CASL probably don't support them on the language level, just when using formal definitions, which say $a := b$ or $a := \neg b$.}
Isabelle, CASL and Mizar allow for defining custom “mixfix” operator notations.  Isabelle provides rich translation mechanisms beyond that, but the layout remains one-dimensional, e.g.\ $\forall x \in A.\; B(x)$ instead of Theorema's $\underset{x\in A}{\forall}\; B[x]$ for bounded quantification.  Isabelle Proof General and Isabelle/jEdit
% support an extensible collection of Unicode glyphs to
approximate textbook notation by Unicode symbols.  Isabelle, Mizar and Hets can export {\LaTeX}.
% \M{} used to support non-ASCII input 
% back in the eighties~\cite{matuszewski2005mizar}
% but then has reverted to plain ASCII.
%% CL@MC: These are interesting details, but I think their essence has been covered in "proof management" already.
% However, as seen in , its language was designed to be a ``simple and intuitive middle-level human-understandable proof language'' \cite{urban2006mizarmode}; with a similar rationale, its chosen proof calculus is a variant (Fitch-Jaskowski, \cite{BancerekR02}) of natural deduction, which is generally considered quite close to how ``ordinary mathematicians construct their proofs'' (\cite{pelletier2012history}, where the interested reader can also find extensive discussion on why this differs from being an effective formal system for other goals of automated reasoning).
\M{} uses ASCII; its lack of binders makes mathematical concepts such as limits and sums cumbersome to denote~\cite{Wiedijk:tqmr07}.
% Shortcomings of its input language include the absence of binders~\cite{Wiedijk:tqmr07}, which makes some fundamental mathematical concepts, such as limits and sums, cumbersome to denote. 
A major reason for us not to cover the TPTP language is its technical, non-extensible ASCII syntax (using, e.g., \lstinline|!|/\allowbreak\lstinline|?| for $\forall$/\allowbreak $\exists$).

Theorema, CASL and \M{} support sharing common quantified variables across multiple statements, corresponding to the practice of starting a textbook section even of several axioms like “let $n$, the number of participants, be a natural number $\ge 1$”.  This helps to avoid redundancy but is prone to copy/\allowbreak paste errors.  For example, our CASL formalisation has sections with global quantifiers $\forall i,j$ (e.g.\ to accommodate the maximum and second-price auction definitions of §\ref{sec:prep-paper-form}), but these include axioms that only use $i$.  Literally pasting into this axiom an expression using $j$ does not cause an error, as $j$ is bound in the current scope as well, but changes the semantics of the axiom in a way hard to detect.
% Similarly, \M{} can reserve a variable name for a specific type; this permits implicit universal quantification over that variable for the scope of the source file.\footnote{{\RaggedRight E.g., \lstinline[language=Mizar]|reserve n for natural number| allows for shortening \lstinline[language=Mizar]|for n being natural number holds n >= 0| to 
% \lstinline[language=Mizar]|n >= 0|.}}

\subsection{Comprehensibility and Trustability of the Output (req.\ \ref{reqC:mrs:understand})}
\label{sec:output}

%\todo{CL@MW: Earlier we had the idea to mention your points about deceptive proofs and irrelevant lemmas here.  There is probably no tool support for them, so I think it would rather be a recommendation w.r.t.\ “general formalisation methodology”, independently of a concrete system? MW: I would say we omit that -- getting too complex.}
Machine proofs may “succeed” for unintended reasons, e.g.\ accidentally stating a tautology such as an implication with an unsatisfiable antecedent.  Or they succeed as intended, but the user cannot follow the (automated) deduction.  In such situations the prover's \emph{output} is crucial.  Isabelle provides tracing facilities for simplification rules and introduction and elimination rules used in standard reasoning steps. %\todo{CL@MW: moved Sledgehammer description to \ref{sec:library-coverage} – OK?}
Its inference kernel can produce a full record (usually large and unreadable) of the internal reasoning of automated tools via explicit proof terms, e.g.\ for independent checking.  By default the kernel relies on static ML type-discipline to achieve correctness by construction, without explicit proof terms.
Theorema's proof %-object
data structure captures the entire proof generation according to the rules and strategy selected. It %is typically
can be displayed as a structured textbook-style proof with configurable verbosity, and visualised as a browsable tree that distinguishes successful from failed branches. 
% \M{} users \todo{CL@MC: Can you phrase this more cautiously?  We haven't done a survey among Mizar users.} normally trust a proof because
\M{} “just” verifies what the user wrote according to natural deduction rules, hence he is unlikely to doubt the result.
On the other hand, for the same reason, \M{} has no way to detect proofs succeeding for unintended reasons, and offers little help to a user clueless about a failing step.  %
%\footnote{It will typically just output error code $4$: \lstinline|This inference is not accepted|.}
A correct \M{} proof can be improved by enhancer utilities~\cite[§4.6]{grabowski2010mizar}: some report useful additional information (e.g., unneeded statements referred in a step, unneeded library files, unneeded lemmas); others cut steps that a human might want to see, impacting readability and possibly the original confidence the user had in the proof.
% However, this might be no longer the case once ancillary proof enhancement tools are employed: they can trim out passages, reducing the obviousness (\todo{CL@MC: rather for my understanding than for elaborating it here: What does \textit{relinfer} do? MC@CL: See source.}\textit{relinfer} is a notable case~).
% Relinfer is used to prune
% "
% unnecessary steps in a proof (so the references may be added to the next step).
% It can exceptionally shorten proofs but it may also result in poorer readability of the text.
% [...] the removal may be accidental in some sense, that is steps which are crucial for human understanding of the idea of a proof, but are still unnecessary for machine (e.g., unwinding definitions-definitional expansions).
% [...] any justification can became (sic) a long list of labels, without significant proof steps which will give the reader an impression about main steps of this proof.
% "
% "
% [...] the relinfer program is dangerous! Not because your
% proofs will break but because they might become ugly. relinfer encourages
% you to cut steps that a human might want to see.
% "
Hets uniformly displays the success of a proof and the list of axioms used; however the latter output is only informative with SPASS.  Otherwise, the raw technical output of the prover is displayed, which strongly differs across provers.  E.g., SPASS uses resolution calculus, which looks different from a textbook proof.  Similarly, System on TPTP outputs performance measures and the status of the given problem %in terms of the SZS ontology
(e.g.\ “Theorem” or “Unsatisfiable”% DROPPED REFERENCE: SZS ontology documentation is linked from the System on TPTP webpage, which is sufficient for us
%~\cite{SZS03}
), but otherwise the raw prover output.

When a proof attempt fails because the statement was wrong, studying a counterexample may help.  Isabelle has the Nitpick counterexample finder built in.  Hets integrates several ones (Darwin is supported best~\cite{Mossakowski:hug06}) and also employs them for consistency checking, as importing a theory whose axioms have no model results in vacuous truth.  %\todo{CL@MW/@TM: FYI I know how, but for reasons of space I'm not saying it.}
Both Isabelle and Hets can attempt a proof or otherwise try to find a counterexample in the same run.  Theorema and Mizar do not support counterexamples.

Before proving, all systems check whether the input is syntactically well-formed and well-typed.  
%\todo{CL@MW: Would you say that Isabelle/jEdit is similarly resilient as Mizar?  I have observed improvements from 2012 to 2013, in that when there is one error, it no longer reports the whole rest of the file as broken.  MW: There is only a slight improvement in 2013; proper structural checking w.r.t.\ sub-proofs as done in Mizar is still missing.}
Isabelle/jEdit performs parsing, type checking and proof processing during editing, and attaches warnings and error messages like modern IDEs.  The other systems require the user to explicitly initiate checking.  % CUT THIS DETAIL: WHAT'S IMPORTANT IS SUBSUMED BY THE PREVIOUS SENTENCE
% Hets' Emacs integration can do just the checks on request, without launching the GUI.
{\M} and Hets check complete files, whereas in Theorema (which only checks syntax), one can individually check each notebook cell (typically containing one to a few statements).
\M's verifier is particularly error resilient: it seldom aborts before the last input line, thus reporting errors for the whole file.

\subsection{Online Community Support and Documentation (req.\ \ref{reqC:mrs:community})}
\label{sec:community-support}

Community support and documentation are major prerequisites for system adoption.  We assume that users with little previous mechanised reasoning and formalisation knowledge %% CL: The following is kind of subsumed by "mechanised reasoning".
% and a system's specific reasoning approach 
will seek low-threshold support from tutorial documents or mailing lists rather than attending community meetings – which, in theorem proving, so far focus on scientific/\allowbreak technical aspects rather than applications.

We compare the community sizes, assuming that large communities are responsive even to non-experts:
Isabelle is developed at multiple institutions% University of Cambridge, Technische Universität München, and Université Paris-Sud
; its user mailing list gets more than 100 posts a month, with over 1000 different authors since 2000% MW: The list has approx.\ 500--1000 subscribers and according to http://groups.google.com/group/fa.isabelle/about it had a bit more than 1000 different authors since the year 2000.
.  CASL, an international standard, has been subject of hundreds of publications % searched Google Scholar for "common algebraic specification language"
but does not currently have a mailing list.  Hets is mainly developed and used within a single institution%Universität Bremen
; its user mailing list receives less than 10 posts a month.  %
% Obviously we didn't count the traffic generated by _this_ collaboration :-)
Recalling that Hets is an integrative environment, users can also request help from the communities of TPTP (subject of more than 1000 publications% searched Google Scholar for tptp (sutcliffe OR "theorem provers")
, no mailing list) and individual provers.  Theorema is developed within a single institution and will not have a mailing list before the 2.0 release.
\M{} is developed at one institution % University of Bia\l{}ystok
by a team % of nine people
that provides dedicated email user assistance: the “\M{} User Service”\todo{CL@MC: temporarily cut the URL footnote: If they did a good job, it should be linked from the homepage anyway.}.  %\footnote{\url{http://megrez.mizar.org/mirror/mus/}}
\mml{}
grows by 30–60 articles a year, with 241 contributors so far% (counted by MML Query~\cite{Bancerek:irrmq06})
.
The mailing list %\todo{CL@MC: This figure is IMHO more relevant than the fact that the developers read mails ;-)}
gets around 10 posts a month.  % long-term average: 1854 posts in the past 180 months (http://mizar.uwb.edu.pl/cgi-bin/wilma/wilma/mizar-forum/archive#browse)

Isabelle and CASL feature comprehensive tutorials and reference manuals, Hets has a user guide, Mizar offers tutorials~\cite{mizar-manuals:on}.  Theorema has \todo{CL@WW: I think that's what I recall from Theorema 1.0}partial built-in help texts and is documented in a few publications.

\newlength{\thw}
\settowidth{\thw}{Documen-}
\newcommand*{\tblhdr}[2][90]{\rotatebox[origin=c]{#1}{~\parbox{\thw}{\baselineskip=.75\baselineskip\RaggedRight #2}}}%\rotatebox[origin=lB]{90}{\fbox{\parbox{\thw}{\RaggedRight #1}}}}
\newcolumntype{R}{>{\centering\arraybackslash}m{.0605\textwidth}}
\ctable[
  width={\textwidth},
  caption={Performance (as far as results were comparable)},
  label={tab:comparison},
  notespar
]{X%
  RRRRRRRRRRRRR%
}{
  \tnote[a]{PI/TI = proof/term input; LC/LS = library coverage/search; PO = proof output; CE = counterexamples (incl.\ consistency checks); WF = well-formedness check.}
  \tnote[b]{scores from very bad (–{}–) to very good (++)}
  \tnote[c]{fully GUI-based}
  \tnote[d]{automated provers}
}{\FL
\multirow{2}{*}{\tblhdr[45]{System/\\ Language}}
             & \tblhdr{Proof\\ speed}
                  & \multicolumn{2}{c}{\tblhdr{Textbook\\ closeness}}
                                & \tblhdr{Top-down\\ proofs}
                                     & \multicolumn{2}{c}{\tblhdr{Library}}
                                               & \multicolumn{3}{c}{\tblhdr{Output}}
                                                                    & \tblhdr{Com\-mu\-nity}
                                                                           & \tblhdr{Documen\-tation}
                                                                                & \tblhdr{de Bruijn\\ factor}
                                                                                  % baseline: TeX 2.6K (2.0K w/o proof)
\NN                 \cmidrule(lr){3-4}
                                       \cmidrule(lr){6-7}
                                                 \cmidrule(lr){8-10}
             &    & PI\tmark[a]
                        & TI\tmark[a]
                                &    & LC\tmark[a]
                                          & LS\tmark[a]
                                                 & PO\tmark[a]
                                                      & CE\tmark[a]
                                                            & WF\tmark[a]
                                                                  &      &    & \NN
             & §\ref{sec:expr-vs.-effic}
                  & §\ref{sec:proof-management}
                        & §\ref{sec:input-syntax}
                                & §\ref{sec:proof-management}
                                     & \multicolumn{2}{c}{§\ref{sec:library-coverage}}
                                                 & \multicolumn{3}{c}{§\ref{sec:output}}
                                                                  & \multicolumn{2}{c}{§\ref{sec:community-support}}
                                                                              & §\ref{sec:level-of-detail} \ML
Isabelle/HOL & ++\tmark[b]
                  & ++  & +     & ++ & ++ & ++   & ○  & ++  & ++  & ++   & ++ & 1.3% 3.5K
\NN
Theorema     & ?  & n/a\tmark[c]
                        & ++    & ++ & +  & –{}– & ++ & n/a & –   & –{}– & –  & n/a%
% was: 3.4; counted by manualy removing CellChangeTimes, "internal cache information" (at the top and bottom of the file), GridBox*
\NN
Mizar        & ++ & ++  & –     & ++ & ++ & +    & ○  & n/a & ++  & +    & ○  & 1.7% 4.4K
\NN
CASL/TPTP    & ○\tmark[d]  & –   & +     & ++ & +  & –    & ○  & +   & +   & ○    & +  & 1.5% 4K
% hets -o pp.stripped.het Vickrey.casl
% tar cf - Vickrey-SecondPriceAuction-*hpf Vickrey-Vickrey-vickreyA.hpf Vickrey.pp.stripped.het | xz -9 >| Vickrey_DeBruijn.xz
\LL}
% MC@CL: Your \M entries are agreeable to me; I added community size based on the number of contributor to the official \mml{}, is that sensible? Is it only a matter of size or also, e.g., of responsiveness? (CL@MC: Yes, responsiveness as well, clarified this – do you consider the community responsive?  The mailing list is not quite as active as Isabelle's.) About prover's output: 
% \M{} verifier's output alone is not so useful, but I kept into account also the ancillary utilities' output. Maybe an external observer has a more detached judging attitude...\\ CL: looks OK to me; not sure how well anyone except you knows Mizar.}

\section{Related Work}
\label{sec:related-work}

§\ref{sec:motivation} mentioned earlier efforts to \emph{formalise economics}.  Particularly Arrow's impossibility theorem, one of the most striking results in theoretical economics, has been a focus for formalisation efforts, including Nipkow's Isabelle and Wiedijk's Mizar formalisation~\cite{nip-09,wie-09}.  As in our case (cf.\ §\ref{sec:prep-paper-form}), they required initial paper elaboration; additionally, it helped them to identify omissions in their source~\cite{gea-01}.  This source states three alternative proofs, but Tang's/\allowbreak Lin's fourth, induction-based proof, allowed for obtaining insights on the general structure of social choice impossibility results using computer support~\cite{ta-li-09}.

The formal verification technique of model checking has been applied to \emph{auctions}.  Tadjouddine et al.\ proved the strategy statement of Vickrey's theorem via two abstractions to reduce the model checker's search space: program slicing to remove variables irrelevant w.r.t.\ the property, and discretising bid values (e.g.\ “higher than someone's valuation $v_i$”)~\cite{ta-gu-va-09}.  Our formalisation is, to the best of our knowledge, the first for \emph{theorem provers}; in the more expressive languages it has the comprehensibility advantage of preserving the structure of the original domain problem.  From earlier economics formalisation efforts cited above, it differs in its goal to (ultimately) help economists to use formal methods themselves.

Our focus thus lies on \emph{comparing} different provers by full parallel formalisation.  Wiedijk compared Isabelle/HOL, Mizar, Theorema, and 14 other provers by general, technical criteria, studying the code resulting from experts formalising a pure mathematics theorem ($\sqrt{2}\notin\mathbb{Q}$), and comparing it to a detailed paper proof~\cite{Wiedijk2006provers}.
% CUT THESE DETAILS:  Where he asked expert users to independently formalise the irrationality of $\sqrt{2}$, we focused on an application domain, and did all formalisations ourselves, influencing each other.  This affords narrower but deeper conclusions.  Wiedijk's “document-centric” comparison focused on “the result of proof formalisation” instead of “the interface of the systems”~\cite{Wiedijk2006provers}.  
%\footnote{In an earlier paper he compares the systems' interaction styles on a high level (proof checking/\allowbreak goal transformation through tactics/\allowbreak automated theorem proving)~\cite{Wiedijk03}.}
We complement this with the end user's perspective: our observations, e.g., on the closeness of the input syntax to textbook notation or the comprehensibility of the output are general, but we emphasised these criteria as they are important to auction designers. 
Griffioen's/\allowbreak Huisman's 1998 PVS and Isabelle/HOL comparison is, like Wiedijk's, independent from a specific application but closer to ours in its look at systems' weaknesses from a user's perspective~\cite{GrifHuis:ComparisonPVSIsabelle98}.  Like us, they rate proof management and user support, but go into more detail up to the “time it takes to fix a bug”.  Their \emph{findings} on user interfaces have been obsoleted by progress in developing textbook-like proof languages and editors with random access and asynchronous validation.\todo{CL: FYI No reviewer requested citing Asperti/Sac.Coen's “Some Considerations on the Usability
of Interactive Provers” (CICM 2010) so we are not going to do it, as we haven't got the space.% Due to its focus on tactics, I didn't find anything useful in there, or can you think of anything? MW: citing is OK for me, although one should point out their main focus on advanced proof automation.\\ CL@MW: If it's merely “OK” I'd say we have no space.  Or would you recommend citing, to demonstrate that we are on topic w.r.t.\ CICM? WW: no opinion\\ CL: answer for now: we haven't got the space.  But I'll keep this note for the revision of the paper, in case a reviewer requests citing.
}

\section{Conclusion and Outlook}
\label{sec:conclusion-outlook}

Auctions allocate trillions of dollars in goods and services every year, but their design is still “far less a science than an art”~\cite{mas-04}.  We aim at making it a science by enabling auction designers to verify their designs.  By parallel formalisation of the first major theorem in a toolbox for basic auction theory (ATT)% and its prerequisites
, we have investigated the suitability of four different theorem provers for this job, taking the perspective not only of experienced formalisers but also of our end users.  Our contribution is 2×2-fold: \begin{enumerate*}\item to auction designers we provide \begin{enumerate*}\item a growing library to build their formalisations on, and \item\label{it:ad-sys} guidelines on what systems to use\end{enumerate*}; \item to the CICM community we provide \begin{enumerate*}\item challenge problems\footnote{Our problems are not currently challenging systems' performance but the promises of their languages and libraries.} and \item\label{it:icm-sys} user experience feedback from a new audience.\end{enumerate*}\end{enumerate*}  This paper focuses on \ref{it:ad-sys} and \ref{it:icm-sys}.

For a concrete application, our findings confirm the widespread intuitions that formalisation benefits from an initial paper elaboration, that the “automated vs.\ interactive” distinction proves of little importance in practice, and that no single system satisfies all requirements%, and this is not due to the “automated vs.\ interactive” distinction
.  \todo{CL@All: We really can't get more concrete than this, can we?  But let's at least once more \emph{explicitly} refer to this table, which kind of is \emph{the} result of this paper.}For now, our comparison results in Tab.~\ref{tab:comparison} guide auction designers in choosing a system, given their formalisation requirements and experience.  The ideal theorem proving environment would feature a \emph{library} as versatile as in Isabelle or Mizar, a \emph{prover} as efficient as those of Isabelle or Mizar, giving \emph{error messages} as informative as in Isabelle/\allowbreak jEdit, further a \emph{proof input language} as close to textbook style as those of Isabelle or Mizar, or an \emph{interface to explore} automated proofs as informative as Theorema's, a \emph{textbook-like term syntax} as Theorema's, an integration of diverse \emph{tools} as in Isabelle or Hets, and a \emph{community} as lively as Isabelle's\todo{CL@All: Rev.\ 2: how do we recommend the user to deal with the systems' weaknesses? – We can't really do anything about this, can we?}.  We have not yet exploited all strengths of the systems evaluated: maintaining a growing ATT with increasingly complex dependencies will benefit from stronger modularisation, as supported by Isabelle and even more so by the theory graph management of Hets/CASL.  Regarding auction \emph{practice}, we are working towards ways to check that formal definitions of auctions are well-defined functions (“for each admissible bid input there is a unique outcome, modulo some randomness”).  Given a constructive proof of this property, it should be possible to obtain verified program code that determines the outcome of an auction given the bids.  This may work using Isabelle's code generator, but we will also explore provers based on constructive type theory\todo{CL: hope we don't want a reference for this…  No space!}.

Broader conclusions about auction theory require further research.  Bidding typically requires forming conjectures of others' beliefs, involving integration over conditional density functions (cf., e.g., Proposition 13 in Maskin's review~\cite{mas-04}).  We expect that much of the required foundations should already be available in the libraries of Isabelle % MW: There seems to be quite a lot of material in HOL-Multivariate\_Analysis, HOL-Probability, leading up to AFP/Markov\_Models, which seems to be the main application here (by J. H\"olzl).
and Mizar% MC: After a cursory look into MML: continuous density functions and sigma fields in PROB\_?, partial derivatives in NDIFF\_5, one-dimensional differentiation in FDIFF\_1, integral calculus (Riemann) in INTEGRA?.
.  Maskin limits his review to single good auctions, noting that few general results exist for multi-unit and combinatorial auctions.\footnote{The last two chapters of~\cite{mil-04} address multi-unit auctions; multi-unit and combinatorial auctions are the focus of~\cite{cr-sh-st-06}.}  Such auctions are often more economically critical (e.g.\ spectrum auctions, monetary policy~\cite{kle-10}) but also more complicated.  \todo{CL@CR: I think here it is not necessary to protect ourselves by starting with “we expect”.}The real challenge for mechanised reasoning will be to demonstrate its use in this domain.\footnote{Even more ambitiously, many results in auction theory are simplified or extended by explicit application of mechanism design; cf.~\cite{Kirkegard:MechanismDesignAuctions2012}.}

% Bib files are in https://codex.cs.bham.ac.uk/svn/langec/formare/lib/bibtex
%TODO MW: to me it looks always a bit odd to see plain website URLs as formal references; one could use footnotes here (in the sense of computer-science footnotes, not social-science ones) -- it will also save pages.
%CL@MW: I'll look into this, but rather closer towards the deadline, once I know how much space we need to save.  Sometimes it does not really save _that_ much.  But as the "footnote area" already exists on most of our pages anyway, I think it will save a bit here. – Independently from space considerations I (being a disciple of Michael Kohlhase) like citing URLs formally, as it allows me to mention title, authors and date.
% \begin{spacing}{0.9}
\printbibliography
% \end{spacing}
\end{document}

%%% Local Variables:
%%% mode: latex
%%% TeX-master: t
%%% LaTeX-biblatex-use-Biber: nil
%%% ispell-local-dictionary: "british"
%%% End:

%  LocalWords:  EP Rowat Kerber Windsteiger Mossakowski RISC Caminati
%  LocalWords:  Hagenberg DFKI GmbH Vickrey's Klemperer CICM ATT FOF
%  LocalWords:  Maskin's Vickrey Maskin Milgrom's alloc maximumExcept
%  LocalWords:  Nipkow's Wiedijk's Geanakoplos Theorema's CASL CASL's
%  LocalWords:  ProofGeneral winningSituation secondPriceAuction AFP
%  LocalWords:  inRange html TPTP's hpf SPASS TFF Makarius Wenzel JKU
%  LocalWords:  Linz Laboratoire LRI UMR Orsay MK's mal thematical de
%  LocalWords:  asoning devVec mizar dBf Bruijn IDE progr plete th LH
%  LocalWords:  CLI errflag addfmsg wrapfigure ISC DoForm LHS xz LZMA
%  LocalWords:  spaWithTruthfulOrOtherBid Isar relinfer SZS hets WFC
%  LocalWords:  notespar RRRRRRRR Wiedijk Griffioen Huisman's PVS wrt
%  LocalWords:  Asperti UsabilityInteractiveProvers subsorts Maeder
%  LocalWords:  renewbibmacro biblatex shortjournal easoning conomics
%  LocalWords:  Milgrom MML Matlab mgmt req MPTP MoMM Miz priori reqs
%  LocalWords:  RRRRRRRRRRRRR WF nity Documen tation lr ers Cramton
%  LocalWords:  ATT's overbar Tadjouddine discretising Griffioen's